\documentclass[a4paper]{article}

\usepackage{INTERSPEECH2019}
\usepackage{xcolor}
\usepackage{mathtools}
\title{Adversarially Trained End-to-end Korean Singing Voice Synthesis System}

\name{Juheon Lee$^{1,2}$, Hyeong-Seok Choi$^1$, Chang-Bin Jeon$^{1,2}$, Junghyun Koo$^1$, Kyogu Lee$^{1,2}$}
\address{
  $^1$Music \& Audio Research Group, Seoul National University\\
  $^2$Center for Super Intelligence, Seoul National University}
\email{\{juheon2, kekepa15, vinyne\}@snu.ac.kr, tonykoo@gmail.com, kglee@snu.ac.kr}

\begin{document}

\maketitle
\begin{abstract}
In this paper, we propose an end-to-end Korean singing voice synthesis system from lyrics and a symbolic melody using the following three novel approaches: 1) phonetic enhancement masking, 2) local conditioning of text and pitch to the super-resolution network, and 3) conditional adversarial training. 
The proposed system consists of two main modules; a mel-synthesis network that generates a mel-spectrogram from the given input information, and a super-resolution network that upsamples the generated mel-spectrogram into a linear-spectrogram.
In the mel-synthesis network, phonetic enhancement masking is applied to generate implicit formant masks solely from the input text, which enables a more accurate phonetic control of singing voice.
In addition, we show that two other proposed methods - local conditioning of text and pitch, and conditional adversarial training - are crucial for a realistic generation of the human singing voice in the super-resolution process.
Finally, both quantitative and qualitative evaluations are conducted, confirming the validity of all proposed methods.
\end{abstract}
\noindent\textbf{Index Terms}: singing voice synthesis, end-to-end network, phonetic enhancement, conditional adversarial training

\section{Introduction} \label{introduction}
With the recent development of deep learning, a learning-based singing voice synthesis (SVS) system, which synthesizes sounds as natural as the concatenative method \cite{macon1997concatenation, bonada2003sample, kenmochi2007vocaloid}, but can expand more flexibly, is proposed. For example, three SVS systems based on DNN, LSTM, and Wavenet architecture were proposed, respectively \cite{nishimura2016singing, kim2018korean, blaauw2017neural}. These systems all include an acoustic model that is trained by singing, lyrics, and sheet music paired data, and each acoustical model is trained to predict the vocoder feature used as an input to the vocoder. 

Although these neural network-based SVS system can achieve adequate performance, networks predicting vocoder features have limits that cannot exceed the upper bound of vocoder performance. Therefore, it is meaningful to propose an end-to-end framework that directly generates a linear-spectrogram, not a vocoder feature. However, the extension to the end-to-end framework of the SVS system is a challenging task because it involves increased complexity of the model. Creating a more complex target, linear-spectrogram, increases the complexity of the model and requires as much training data to generalize and train these models sufficiently. However, gathering singing audio with aligned lyrics in a controlled environment is a task that requires a lot of effort. 

We proposed in this paper a Korean SVS system that can be trained by an end-to-end manner with moderate amounts of data \footnote{The generated result can be found at: ksinging.strikingly.com.}. 
Our baseline network is designed with the inspiration of DCTTS \cite{tachibana2018efficiently}, known as efficiently trainable text to speech (TTS) system. We applied the following novel approaches to enable end-to-end network training. First, we used the phonetic enhancement masking method, which separately modeled low-level acoustic features related to pronunciation from text information, to make more efficient use of the information contained in the training data. Second, we also proposed a method of reusing input data at the super-resolution stage and training with an adversarial manner to produce better sound quality singing. 

The contribution of this paper is as follows: ${\textbf{1)}}$ We designed the end-to-end Korean SVS system and suggested a way to train it effectively. ${\textbf{2)}}$ We proposed a phonetic enhancement masking method that helps to produce more accurate pronunciation. ${\textbf{3)}}$ We proposed a conditional adversarial training method for the generation of more realistic singing voices. \\

\begin{figure*}[ht]
\centering
\includegraphics[width=0.9\linewidth]{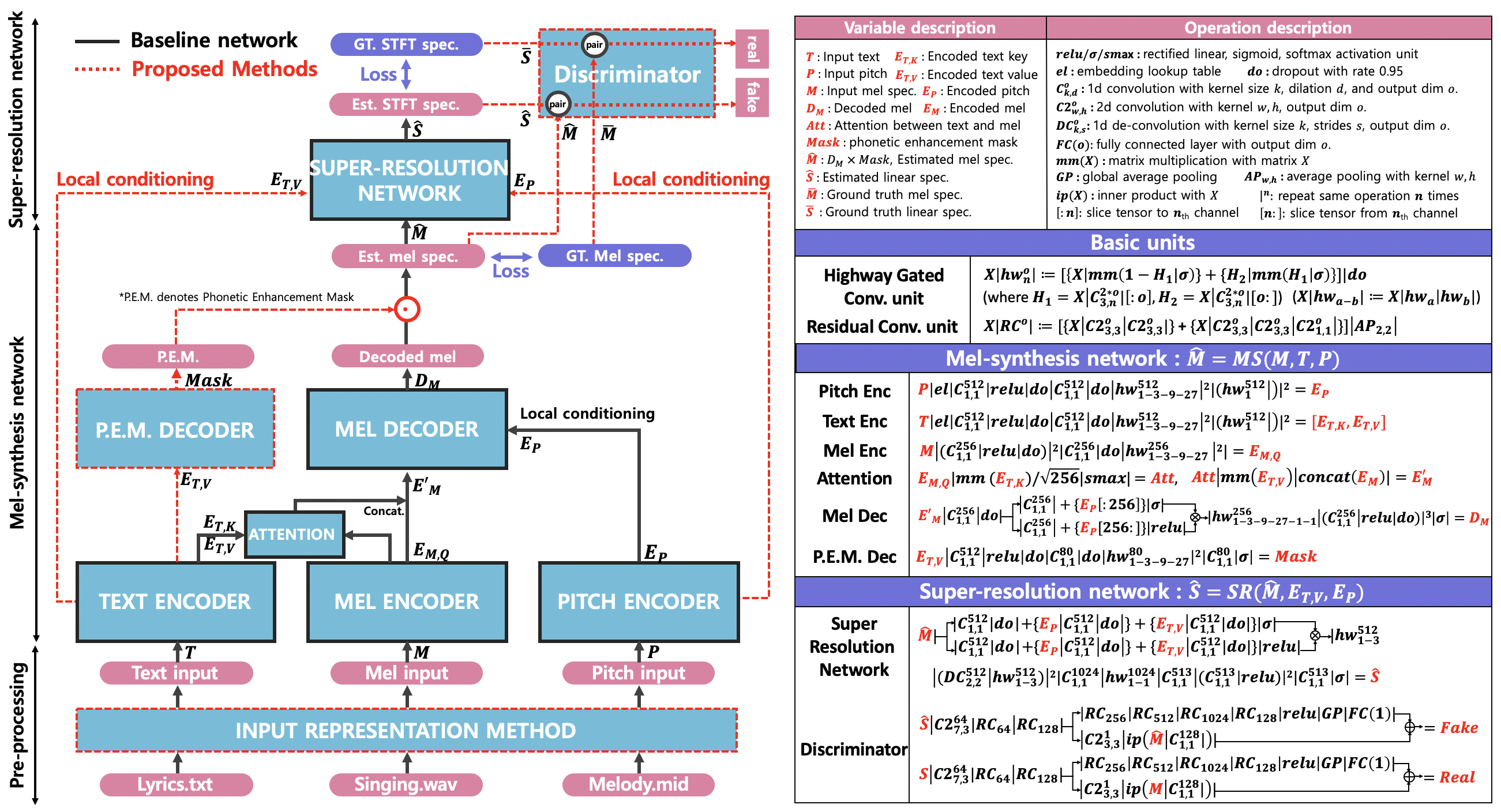}
\caption{
Proposed system overview (left), detailed structure of each sub-module (right). $X|F|$ denotes $F(X)$.
} 
\label{fig:network}
\end{figure*}


\section{Related work} \label{related}
The SVS system is similar to the TTS system in terms of synthesizing natural human speech. Recently, the end-to-end TTS system, which is trained as an autoregressive manner, such as Tacotron\cite{wang2017tacotron}, Deep voice\cite{gibiansky2017deep}, is showing better performance than the conventional method. In addition, various follow-up studies are being conducted that have further controllable elements such as prosody, style, etc \cite{skerry2018towards, wang2018style}, or models that can be trained more efficiently \cite{tachibana2018efficiently, chung2018semi}. We conducted the study by modifyng the TTS model to suit the SVS task, based on DCTTS\cite{tachibana2018efficiently}, which is known to be capable of efficient end-to-end training.

The generative adversarial networks (GAN) is a widely used technique that helps train an arbitrary function to generate a similar sample as the sample from desired data distribution. This training method has been widely accepted in computer vision community and becomes one of the key components to attain photo-realism in super-resolution task. Unlike the success of adversarial training method in image domain, however, only a few works have achieved a reasonable success of training super-resolution task (specifically, band-width extension task) in audio signal processing community \cite{li2018speech}. To further leverage the promise of adversarial training in audio generation process, we adopted a few recent works that stabilizes the adversarial training, namely, conditional GAN with projection discriminator \cite{miyato2018cgans} and R1 regularization \cite{brock2018large} which allow us to jointly train the autoregressive network (mel-synthesis) and super-resolution network making the proposed system as an end-to-end framework.


\section{Proposed Network} \label{proposed}
As illustrated in Figure \ref{fig:network}, our proposed model consists of two main modules, a mel-synthesis network and a super-resolution network. The mel-synthesis network is trained to produce a mel-spectrogram $M_{1:L}$ from previous mel input $M_{0:L-1}$, time-aligned text $T_{1:L}$, and pitch inputs $P_{1:L}$. With text and pitch information as conditional input, the super-resolution network upsamples the generated mel-spectrogram $M$ to a linear-spectrogram $S$. Finally, the discriminator takes the upsampled result with generated mel-spectrogram to train the network in an adversarial manner. 
During the test phase, a sequence of mel-spectrogram frames is generated in an autoregressive manner from a given text and pitch input which is then upsampled to linear-spectrogram by super resolution network. Finally, the generated linear-spectrogram is converted to a waveform using Griffin-Lim algorithm \cite{griffin1984signal}.

\begin{figure}[h]
\centering
\includegraphics[width=0.9\linewidth]{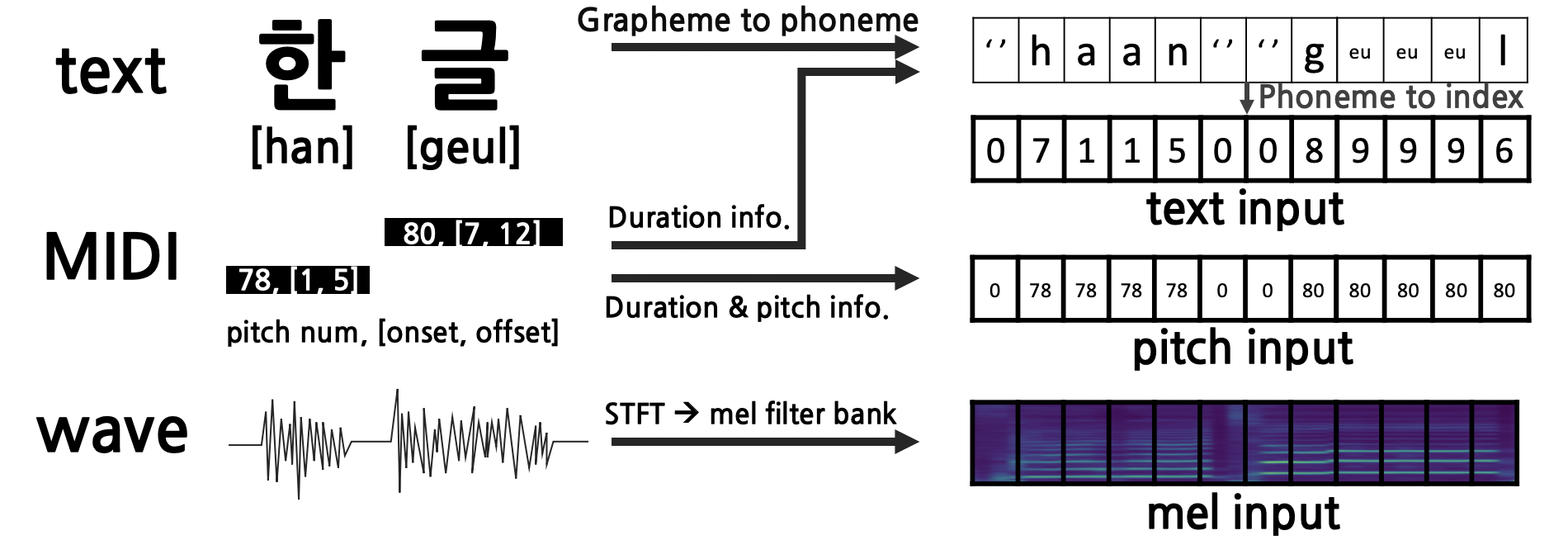}
\caption{
Input representation method overview} 
\label{fig:input}
\end{figure}

\subsection{Input representation} \label{input}
Our training data includes recorded singing voice along with the corresponding text and midi. A single midi note represents pitch information with onset and offset. For the single midi note, one syllable and its corresponding vocal audio section are manually aligned. 
Figure \ref{fig:input} shows our input representation more concretely.
To determine the text input sequence $T \in {\mathbb{R}}^{1 \times L}$ with length $L$, we referred to the pronunciation system of Korean. A Korean syllable can be decomposed into three phonemes each of which corresponds to onset, nucleus, and coda, respectively.
Since the nucleus occupies most of the pronunciation singing in Korean, we assigned onset and coda to the first and the last frame of input text array, respectively, and the rest of the frames with nucleus.
Although this does not reflect accurate timing for each phoneme, we empirically found out that a convolution-based network with wide enough range of receptive field can handle this problem.
For pitch input $P \in {\mathbb{R}}^{1 \times L}$ , we simply assigned a pitch number to each frame. 
In the case of the mel input $M \in {\mathbb{R}}^{F \times L}$, we used the mel-spectrogram itself, which was extracted from the recorded audio, where $F$ denotes the number of frequency bins.

\subsection{Mel-synthesis network} \label{mel-synthesis}

The mel-synthesis network $MS(\cdot{})$ aims to generate the mel-spectrogram of the next time step from the given text, pitch, and mel input. 
Based on the text-to-mel network proposed by \cite{tachibana2018efficiently} we modified it to fit the SVS system. 

First, in order to enter pitch information, we added pitch encoders with the same structure as text encoders. In addition, the local conditioning method proposed by \cite{oord2016wavenet} was used to conduct a conditioning of the encoded pitch on the mel decoder. 

Second, we assumed that among the various elements forming a singing voice, information about pronunciation would be able to be controlled independently from text information. We also assumed that if the low-level audio feature that constitutes pronunciation information can be modeled independently, it is possible to focus on the pronunciation information in the data composed of various combinations of pronunciation-pitch, so that training data can be utilized more efficiently to generate more accurate pronounced singing voice. To this end, we designed an additional phonetic enhancement mask decoder, which receives encoded text only as input, and the output of the decoder element-wise multiplied by the output of the mel decoder to create the final mel-spectrogram. As a result, $MS(\cdot)$ can be formulated as follows:

\begin{equation}
\hat{M} = Mask \odot D_M = MS(M,T,P)
\end{equation}

We trained the $MS(\cdot)$ network with $L_1$ and binary divergence loss $L_d$ between ground truth and generated mel-spectrogram, and guided attention loss $L_{att}$ as the objective function. Please see \cite{tachibana2018efficiently} for more detailed explanation on the loss terms.
We also assumed that the $L_1$ loss between the differential spectrogram $M' = M_{1:L} - M_{0:L-1}$ would be also beneficial for network to learn more about the relatively short pronounced onset, coda. Therefore, the overall objective function for $MS(\cdot)$ is as follows:
\begin{equation}
    L_{MS} = L_1(\hat{M}, M) + L_d(\hat{M}, M) + L_{att} + L_1(\hat{M'},M')
\end{equation}

\subsection{Super-resolution network} \label{super-resolution}
In this section, we describe the details of the training method for super-resolution network $SR(\cdot)$.
The purpose of the SR step is to upsample the generated mel-spectrogram $\hat{M} \in {\mathbb{R}}^{F \times L} $ into a linear-spectrogram $\hat{S} \in{\mathbb{R}}^{F' \times L'}$ thereby making it to an audible form, where $F'$ and $L'$ denote the number of frequency bins and temporal bins for linear-spectrogram.
The idea of the SR network was proposed in a few previous TTS literatures, including Tacotron and its variants \cite{tachibana2018efficiently, wang2017tacotron}.
The major difference between the previous works and our work is twofold. 
First, we additionally reuse the aligned text and pitch information into the SR network exploiting the useful information in the generation process again.
Second, we utilize adversarial training methods to make the SR network produce more realistic sound.

\subsubsection{Local conditioning of text and pitch information} \label{local}
Unlike the attention-mechanism based TTS literature, SVS system requires the aligned text and pitch information as inputs for the controllability in the generation process. These information, therefore, can be easily reused in the SR step in the absence of time-alignment process as follows.
\begin{equation}
\hat{S} = SR(\hat{M}, E_{T,V}, E_P) = SR(MS(\cdot), E_{T,V}, E_P)
\end{equation}
More specifically, each of the output from the text encoder and pitch encoder ($E_{T,V}$ and $E_P$) is fed into a sequence of $1\times1$ convolutional and dropout layer \cite{srivastava2014dropout} which is then fed into a highway network as a local conditioning method as proposed in \cite{oord2016wavenet}. For the upsampled $\hat{S}$, SR is trained with the objective function $L_{SR} = L_1(\hat{S},S) + L_d(\hat{S},S)$. For the exact network configuration, please refer to Figure \ref{fig:network}.

\subsubsection{Adversarial training method} \label{adversarial}
Expecting to generate a realistic sound, we adopted a conditional adversarial training method which helps the output distribution of $\hat{S}=SR(\hat{M}, \cdot)$ be similar to the real data distribution $S \sim p(S|M)$.
Intuitively, in the conditional adversarial training framework, discriminator $D_{\psi}$ not only tries to check if $S$ is realistic but also the paired correspondence between $S$ and $M$.
Note that, we make a minor assumption that the distribution of $\hat{M}=MS(\cdot)$ approximately follows that of $M$, that is, $p(M)\simeq p(\hat{M})$, allowing the joint training of two modules $MS(\cdot)$ and $SR(\cdot)$.
The conditioning to discriminator was done by following \cite{miyato2018cgans} with a minor modification. First, the condition $M$ is fed into a 1d-convolutional layer and the intermediate output of discriminator is fed into a $3\times3$ 2d-convolutional layer. Then, inner product between the two outputs is done as a projection. Finally, the obtained scalar value is added to the last layer of $D_{\psi}$ resulting in final logit value. For the exact network configuration please refer to Figure \ref{fig:network}.

For the stable adversarial training, a regularization technique on $D_{\psi}$ has been proposed by several GAN related works \cite{gulrajani2017improved, miyato2018spectral, mescheder2018training, brock2018large}. We adopted a simple, yet, effective gradient penalty technique called R1 regularization. This technique penalizes the squared 2-norm of the gradients of $D_{\psi}$ only when the sample from true distribution is taken as follows
\begin{equation}
R_1(\psi)=\frac{\gamma}{2} \mathbb{E}_{p(M,S)}[\|\nabla D_{\psi}(M,S)\|^2].
\end{equation}
Note that the output of $D_{\psi}$ denotes the logit value before the sigmoid function.
The final adversarial loss terms ($L_{adv_D}$ and $L_{adv_{G}}$) for $D_{\psi}$ and $G_{\theta}$ are as follows,
\begin{equation}
\label{adv_loss}
\begin{split}
L_{adv_D}(\theta, \psi) &= -\mathbb{E}_{p(M)}[\mathbb{E}_{p(S|M)}[f(D_{\psi}(M,S))]] \\
& - \mathbb{E}_{p(\hat{M})}[f(D_{\psi}((\hat{M},\hat{S}))] + R_1, \\
L_{adv_{G}}(\theta, \psi) &= \mathbb{E}_{p(\hat{M})}[f(D_{\psi}(\hat{M},\hat{S}))],
\end{split}
\end{equation}
where $\theta$ includes not only the parameters of $SR$ but also that of $MS$, hence the two consecutive modules acting as one generator function $G_{\theta}=SR(MS(\cdot),\cdot)$. The function $f$ is chosen as follows $f(t)=-log(1+exp(-t))$ resulting in the vanilla GAN loss as in the original GAN paper \cite{goodfellow2014generative}.

\section{Experiments} \label{experiments}
\subsection{Dataset}

Since there is no publicly available Korean singing voice dataset, we created the dataset as follows.
First, we prepared accompaniment and singing voice MIDI files of 60 Korean pop songs. Next, a professional female vocalist was told to sing to the accompaniment. Then, the singing voice MIDI files were manually realigned so that the recorded audio have the exact alignment with the singing voice MIDI files.
Finally, we manually assigned the syllables in lyrics to each MIDI note of singing voice MIDI file.
The audio length of the entire dataset excluding the silence is about 2 hours. We used 49 songs for training dataset, 1 song for validation, and 10 songs for test dataset.

\subsection{Training} \label{training}
We trained the discriminator to minimize $L_{adv_D}$ and the rest of the network to minimize $L_{adv_G}$, $L_{MS}$ and $L_{SR}$. For SR networks, we have to start training after the appropriate level of mel is generated, so we have separately controlled $lr_{SR}$ and $lr_{GAN}$ to add to the objective function. At this point, it was set to $lr_{SR} = \rm{min}(0.2*(iter/100), 1)$, $lr_{GAN} = \rm{min}(0.01*(\rm{int})(iter/5000), 1)$, respectively.
\begin{equation}
\begin{split}
    L_{MS, SR} &= L_{MS} + lr_{SR} \cdot L_{SR} + lr_{GAN} \cdot L_{adv_G} \\
    L_D &= lr_{GAN} \cdot L_{adv_D}
\end{split}
\end{equation}
In both cases, we used Adam optimizer \cite{kingma2014adam}, which was set to $\beta_1$ = 0.5 and $\beta_2$ = 0.9. The learning rate was scheduled to start from 0.0002 and was halved for every 30,000 iteration. All parameters of the networks were initialized with the Xavier initializer \cite{glorot2010understanding}.

For the ground truth mel/linear-spectrogram, we first extracted the linear-spectrogram $S$ from audio with $sr=22050, n_{fft}=1024, hop=256$. 
We then normalized the linear-spectrogram as follows $S \leftarrow (|S|/max(|S|))^\delta$, where $\delta$ denotes a pre-emphasis factor with the value of 0.6 in our case. \footnote{Note that we post emphasized $\hat{S} \leftarrow \hat{S}^{\zeta/\delta}$ where $\zeta $ denotes a post-emphasis factor with the value of 1.3.}
Afterwards, the mel-spectrogram was obtained by multiplying 80-d of mel filter bank to $S$, and the same normalization method as in $S$ was used. In order to reduce the complexity of the model, we downsampled the mel-spectrogram to the quarter by taking the first frame of every four-frame of the mel-spectrogram giving the relationship of $L' = 4L$.



\subsection{Evaluation} \label{evaluation}
We trained a total of five models to see how the three proposed methods - method 1; phonetic enhancement masking, method 2; local conditioning pitch and text to $SR(\cdot)$, method 3; adversarial training method - actually affect the network. 
The differences between the five models are described in Table 1. 
20 audio samples from each model were generated from the test dataset.
Apart from the generated samples, we also compared the ground truth samples.
\textbf{Ground} denotes the actual recorded audio, and \textbf{Recons} denotes the reconstructed audio from ground truth magnitude only linear-spectrogram using Griffin-Lim algorithm. 
Noe that \textbf{Recons} samples were included to evaluate the sound quality from the loss of phase information.


\begin{table}[h]

\caption{ Model description and evaluation result.}
\centering
\scalebox{0.7}{
\setlength\tabcolsep{4pt}
\begin{tabular}{c|c|c|c|c|c}
\toprule

\textbf{Model}  &  model1 & model2 & model3 & model4 & model5 \\ \hline
\textbf{Method} &  baseline & +(method 1) & +(method 2) & +(method 1,2) & +(method 1,2,3)\\

\bottomrule
\end{tabular}
}

\centering
\scalebox{0.7}{
\setlength\tabcolsep{4pt}
\begin{tabular}{c|ccc||cccc}
\toprule

& \multicolumn{3}{|c||}{\textbf{Quantitative}} & \multicolumn{3}{|c}{\textbf{Qualititative}} \\ \hline
Model      & Precision     & Recall      & F1-score & Pronun.acc & Sound.quality & Naturalness \\ \hline
model1     & 0.771 & 0.832  & 0.800   & $2.29 \pm 1.15$ & $2.32 \pm 0.89$ & $2.11 \pm 0.98$ \\
model2     & 0.780 & \textbf{0.843}  & 0.810  & $2.62 \pm 1.00$ & $2.28 \pm 0.84$ & $2.22 \pm 0.91$ \\
model3     & 0.755 & 0.814  & 0.783  & $2.69 \pm 1.06$ & $2.37 \pm 0.86$ & $2.22 \pm 0.93$ \\
model4     & 0.792 & 0.832  & 0.811  & $2.92 \pm 1.08$ & $2.43 \pm 0.86$ & $2.36 \pm 0.94$ \\
model5     & \textbf{0.872} & 0.821  & \textbf{0.846}  & $\textbf{3.23} \pm \textbf{1.19}$ & $\textbf{3.37} \pm \textbf{0.94}$ & $\textbf{3.07} \pm \textbf{1.10}$ \\ \hline
Recons     & 0.805 & 0.830  & 0.782  & $4.85 \pm 0.47$ & $4.46 \pm 0.77$ & $4.72 \pm 0.62$ \\
Ground     & 0.826 & 0.772  & 0.798  & $4.90 \pm 0.36$ & $4.74 \pm 0.57$ & $4.85 \pm 0.43$ \\

\bottomrule
\end{tabular}
}
\end{table}


\subsubsection{Quantitative evaluation} \label{quantitative}
We evaluated whether the network was actually producing a conditioned singing voice for a given input. To do this, we extracted f0 sequence from the generated audio through the world vocoder\cite{morise2016world}, converted it into a pitch sequence, and compared it to the input pitch sequence. We can judge that the higher the similarity between the two sequence, the more the network generates a singing that reflects the input condition. We calculated the precision, recall and f-score of the generated pitch sequence by frame-wise, and the results are shown in Table 1. 

Even in the case of a real recording sample recorded by listening to the original midi accompaniment, it is not easy to adjust the timing and pitch of the correct note, so that a 100\% accurate f-score can not be obtained. For all samples that were generated, a f-score similar to or higher than the real recording sample was obtained. This means that the model has generated a singing voice with the correct pitch and timing for at least the real recording for the given input.

\subsubsection{Qualitative evaluation} \label{qualitative}
We conducted a listening test to evaluate the quality of the generated singing voice.
19 native Korean speakers were asked to listen to the 20 audio samples from each model.
Each participant was asked to evaluate the pronunciation accuracy, sound quality, and naturalness.
During the listening test, lyrics of audio samples were provided for more accurate evaluation of pronunciation accuracy.
The MOS results are shown in Table 1.

We conducted a paired t-test for each model response and based on this we verified the effectiveness of the proposed methods. 
For the accuracy of the pronunciation, we obtained significant differences for all comparisons except for models 2 and 3. 
In other words, all of the proposed methods helped to create more accurate pronunciation singing voices, and the performance was improved to the greatest extent with all three methods. 
In the case of sound quality, methods 1 and 2 did not significantly affect the improvement, but the applying method 3 showed a significant increase in score.  
From this we can confirm that training the network in an adversarial manner improves the quality of the generated audio. 
Finally, for naturalness, there was a significant improvement when all methods were applied.

\subsection{Analysis on generated spectrogram} \label{analysis}



\begin{figure}[t]
\centering
\includegraphics[width=0.9\linewidth]{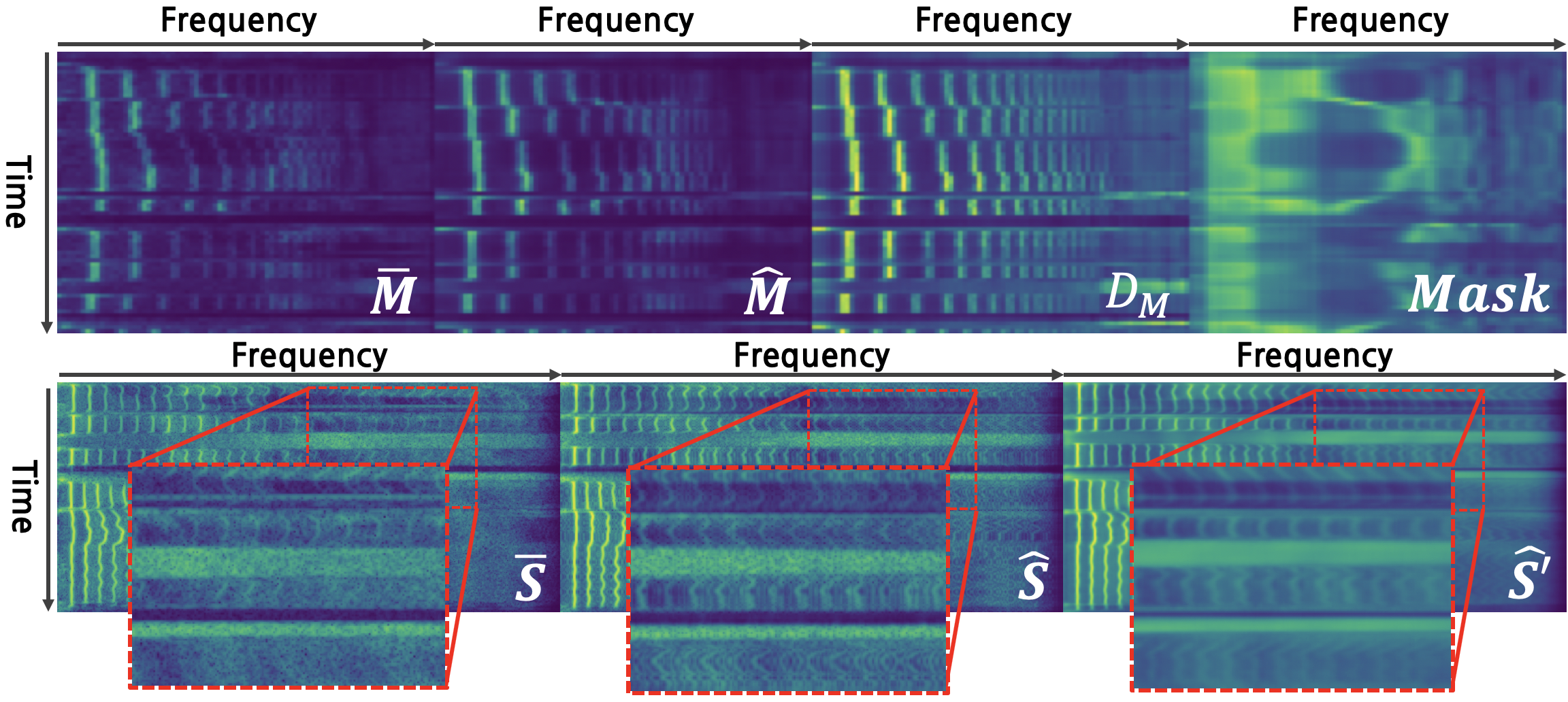}
\caption{
Generated spectrograms.} 
\label{fig:analysis}
\end{figure}

In this section we analyze the features generated by the mel-synthesis and super-resolution networks. In the case of mel-synthesis network, from observing internally generated features, we found that the low-level acoustic feature of pronunciation and pitch could be divided independently without any supervision. From Figure \ref{fig:analysis}, $D_M$ shows the underlying structure of the spectrogram, such as the harmonic structure and the location of f0. In $Mask$, on the other hand, we can observe the shape of determining the intensity of the frequency at every time-step, similar to the feature of the spectral envelope, which contains non-periodic information. This suggests that, from the perspective of source-filter models, one of the techniques that classical speech modelling techniques, our network can generate sources ($D_M$) and filters ($Mask$) separately from frequency domain without any supervised training.

We also analyzed the effect of adversarial training method by observing the generated linear-spectrogram. 
Three different spectrograms from model4 ($\hat{S'}$: w/o adversarial loss), model5 ($\hat{S}$: w/ adversarial loss), and ground truth spectrogram ($\bar{S}$) are demonstrated in the second row of Figure \ref{fig:analysis}.
While $\hat{S'}$ showing the blurry high frequency areas, $\hat{S}$ clearly shows that adversarial training allows the proposed network to generate sample that is closer to the ground truth sample $\bar{S}$.
Note that we have confirmed in \ref{qualitative}, listening test that the sound quality can be significantly improved by comparing model4 and model5, which again reinforces our observation.

\section{Conclusions}
In this paper, we proposed the end-to-end Korean singing vocie synthesis system. We showed that using text information to model the phonetic enhancement mask actually worked, and produced more accurate pronunciation. Also, we successfully applied the conditional adversarial training method to the super-resolution stage, which resulted in a higher quality voice.

\section{Acknowledgements}
This work has partly supported by National Research Foundation of Korea (NRF) funded by the Korea government (NRF-2017R1E1A1A01076284), and partly by Institute for Information \& Communications Technology Planning \& Evaluation(IITP) grant funded by the Korea government (No.2019-0-01367)

\bibliographystyle{IEEEtran}

\bibliography{mybib}


\end{document}